\documentclass[aps,prx,reprint,groupedaddress,amsmath,amssymb,floatfix]{revtex4-2}
\usepackage{graphicx}
\usepackage{bm}
\usepackage{booktabs}

\usepackage{hyperref}
\hypersetup{colorlinks=true,citecolor=blue,linkcolor=blue,urlcolor=blue}
\AddToHook{env/thebibliography/begin}{\interlinepenalty=10000\relax}
\graphicspath{{figures/}}
\newcommand{\Ca}[1]{^{#1}\mathrm{Ca}}

\begin{document}
\title{Learning nuclear mass systematics with hybrid quantum–classical mixtures of experts}
\author{Shijiang Ji}
\thanks{These authors contributed equally to this work.}
\affiliation{School of Physical Science and Technology, Southwest University, Chongqing 400715, China}

\author{Heyi Yang}
\thanks{These authors contributed equally to this work.}
\affiliation{School of Physical Science and Technology, Southwest University, Chongqing 400715, China}

\author{Zu-Xing Yang}
\thanks{yangzuxing@swu.edu.cn.}
\affiliation{School of Physical Science and Technology, Southwest University, Chongqing 400715, China}

\author{Xiao-Hua Fan}
\affiliation{School of Physical Science and Technology, Southwest University, Chongqing 400715, China}

\author{Zhi-Pan Li}
\affiliation{School of Physical Science and Technology, Southwest University, Chongqing 400715, China}

\author{Haozhao Liang}
\affiliation{Department of Physics, Graduate School of Science, The University of Tokyo, Tokyo 113-0033, Japan}
\affiliation{RIKEN Interdisciplinary Theoretical and Mathematical Sciences Program, Wako 351-0198, Japan}

\begin{abstract}
We introduce a mixture-of-experts framework combining sixteen hybrid quantum–classical neural networks with nucleus-dependent classical gating to learn residual corrections to the WS4 nuclear mass model. 
On a fixed AME2020 test set, the framework reduces the root-mean-square error from 456.22 to 179.83 keV and outperforms a parameter-matched classical counterpart.
The learned corrections improve shell-gap systematics near the Sn and Pb shell closures, while analysis of the gating networks identifies shell-related descriptors as the leading inputs. 
For \(^{66}\mathrm{Ca}\), the model predicts \(S_n=0.577\) MeV and \(S_{2n}=-0.252\) MeV, allowing two-neutron emission while the one-neutron channel remains closed.
\end{abstract}
\maketitle
\raggedbottom

\section{Introduction}

Nuclear binding energies are fundamental to understanding nuclear structure and stability, as well as the energetics of nuclear reactions and decays. 
Their systematics across the nuclear chart reflect the interplay of bulk properties, shell structure, and pairing correlations. 
Differences between neighboring masses provide sensitive probes of shell evolution and determine particle-emission thresholds.
Accurately capturing both global mass trends and the corresponding local variations remains a central challenge for nuclear mass models, particularly when extending predictions far from stability.

The Weizs\"acker--Skyrme model (WS4) provides an accurate macroscopic--microscopic description of nuclear masses by combining global trends, microscopic corrections, and a surface-diffuseness correction~\cite{wang2014ws4}.
Alongside physics-based models, data-driven approaches have long been used to learn nuclear mass systematics directly from data~\cite{athanassopoulos2004}.
Residual learning connects these approaches by retaining a physical mass model as the baseline and learning corrections to its deviations from reference masses. 
Within this framework, radial-basis-function corrections capture spatial patterns in these residuals~\cite{wang2011rbf}, while Bayesian neural networks refine existing mass formulas and provide uncertainty estimates~\cite{utama2016bnn,utama2017}. 
Their predictive power can be further enhanced by incorporating pairing and shell descriptors~\cite{niu2018}, and learning an even–even mass surface jointly with neighboring-nucleus correlations offers another route to encode nuclear structure~\cite{niu2022nuclear}. 
Probabilistic mixture-density networks extend this direction by predicting full mass distributions~\cite{lovell2022}, and physically interpretable networks integrate nuclear descriptors with soft physics constraints~\cite{mumpower2022}. 
Even multilayer feedforward networks have been applied within a residual-learning setting, improving liquid-drop residual predictions and separation-energy trends~\cite{le2023}. 
In our previous work, we used multitask ensemble learning to connect nuclear densities, charge radii, and binding energies in a data-driven density-functional model~\cite{yang2024dddf}.
Related subsequent work explored mirror-nucleus mass--radius correlations~\cite{alam2025mirror} and XGBoost residual corrections with nuclear-structure descriptors~\cite{zhang2026xgboost} for nuclear mass prediction.

A hybrid quantum–classical neural network (HQNN) first encodes physical descriptors through classical layers, passes them through a parameterized quantum circuit, and maps the measured expectation values to a prediction~\cite{mitarai2018qcl,cerezo2021vqa}. 
Because the circuit evolves in a high-dimensional Hilbert space, this representation can explore rich nonlinear feature spaces with a comparatively small number of trainable parameters; the choice of data encoding further shapes the function class accessible to the model~\cite{schuld2021encoding}.
Besides, where a single expert may struggle to cover the whole nuclear chart uniformly, a mixture-of-experts (MoE) architecture takes a complementary route: it pools multiple specialized experts and learns nucleus-dependent gating weights~\cite{jacobs1991moe}, so that different expert combinations are selected across the chart. 
Originally proposed for general regression and classification, MoE has since become a standard scalability mechanism in deep learning. 
It underpins large language models such as Switch Transformer~\cite{fedus2022switch}, GShard~\cite{lepikhin2021gshard}, Mixtral~\cite{jiang2024mixtral}, DeepSeekMoE~\cite{dai2024deepseekmoe}, Qwen2-MoE~\cite{yang2024qwen2}, and Grok-1~\cite{xai2024grok}.
Sparse expert activation allows model capacity to grow without a proportional increase in computation per input~\cite{fedus2022switch,lepikhin2021gshard}.
MoE architectures have also been adopted in computer vision~\cite{riquelme2021vision} and computational-physics surrogates, including gravitational-waveform models~\cite{schmidt2021mlgw}.

In this work, we combine sixteen HQNN residual experts with nucleus-dependent classical gating to refine the WS4 mass surface. 
Comparisons with parameter-matched classical experts and uniform expert averaging allow us to examine the roles of the hybrid representation and adaptive weighting. 
We assess global mass accuracy and analyze the learned corrections through the gating networks' use of nuclear-structure descriptors and local mass differences sensitive to shell structure. 
We then examine the predicted drip lines and use neutron-rich calcium to resolve their implications for individual neutron-emission thresholds.
The remainder of this paper is organized as follows. 
Section II introduces the HQNN and artificial neural network (ANN) residual models and the MoE framework. 
Section III evaluates global mass accuracy and analyzes the role of nuclear-structure descriptors in expert weighting.
It then examines separation energies and higher-order mass differences along the Ca, Ni, Sn, and Pb isotope chains. 
Section IV connects the spatial mass residuals to predicted two-nucleon drip lines, followed by a closer analysis of neutron-rich calcium. 
Section V summarizes the main findings.

\section{Residual models and mixtures of experts}
For a nucleus with proton number $Z$ and neutron number $N$, we define the residual $r(Z,N)$ as
\begin{equation}
r(Z,N)=B_{\mathrm{WS4}}(Z,N)-B_{\mathrm{exp.}}(Z,N),
\label{eq:targetA}
\end{equation}
where $B$ denotes the total binding energy.
WS4 is adopted as the theoretical baseline, and the experimental data are taken from the Atomic Mass Evaluation 2020 (AME2020)~\cite{wang2021ame}.

\subsection{Hybrid quantum–classical neural network}

\begin{figure}[tb]
\centering
\includegraphics[width=0.45\textwidth]{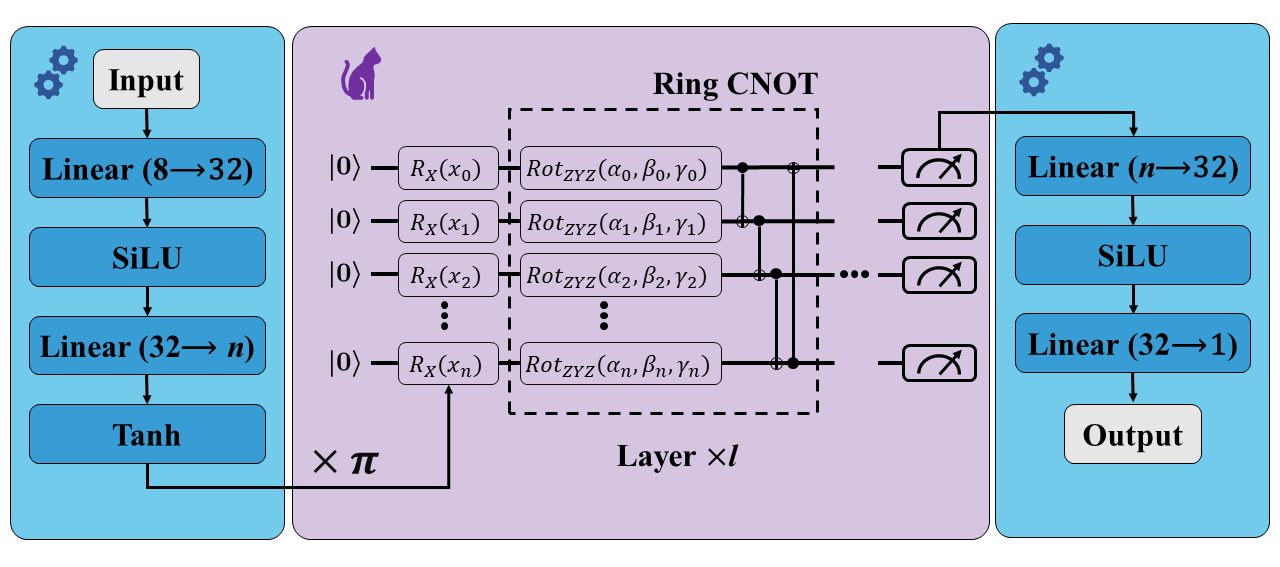}
\caption{Architecture of a hybrid quantum–classical neural network (HQNN) residual expert. 
A classical encoder maps eight nuclear descriptors to $n$ encoding angles, and a classical readout maps $n$ Pauli-$Z$ expectation values to the residual. 
See the text for details of the quantum circuit.
}
\label{fig:main1}
\end{figure}

In this work, the HQNN serves as the primary expert model. 
As shown in Fig.~\ref{fig:main1}, it follows the general structure of a classical neural network, comprising input, hidden, and output layers. 
Its input combines the proton and neutron numbers with six additional features to form the eight-dimensional feature vector
\begin{equation}
\bm{u}=\{Z,N,d_Z,d_N,N/Z,\delta,P,A\},
\label{eq:expertfeatures}
\end{equation}
where $d_Z$ and $d_N$ are distances to the nearest conventional magic number in $\{2,8,20,28,50,82,126\}$ and $P=+1$, $-1$, and 0 for even--even, odd--odd, and odd--A nuclei, respectively. 
The parameter $\delta=(N-Z)/A$ encodes the neutron--proton asymmetry, where $A=N+Z$ is the mass number.
The input vector $\bm{u}$ is Z-score-standardized to $\widetilde{\bm u}$.
Hereafter, a tilde denotes standardized quantities.

Each HQNN expert consists of a classical encoder, a parameterized quantum circuit, and a classical readout.
The encoder converts the nuclear descriptors into rotation angles, the circuit transforms the encoded state, and the readout converts measured expectation values into a residual prediction. 
The encoder has widths $8\to32\to n$ and produces
\begin{equation}
\bm\theta = \pi\tanh(W_2 \operatorname{SiLU}(W_1\widetilde{\bm u}+\bm b_1)+\bm b_2),
\label{eq:hqnnencoder}
\end{equation}
where $\bm\theta=(\theta_1,\ldots,\theta_n)$ contains the encoding angles. 
Here, $W_1,W_2,$ and $\bm b_1,\bm b_2$ are trainable weights and biases, respectively. 
The SiLU activation is $a/(1+e^{-a})$ and the final $\tanh$ bounds each angle to $(-\pi,\pi)$.

These angles are encoded through single-qubit $X$ rotations on the initial state $|\bm 0\rangle\equiv|0\rangle^{\otimes n}$:
\begin{equation}
U(\bm\theta)=\bigotimes_{\ell=1}^{n}R_x(\theta_\ell),~~~
|\psi_{\mathrm{enc}}\rangle=U(\bm\theta)|\bm 0\rangle.
\label{eq:hqnnencoding}
\end{equation}
We use the convention $R_\nu(\vartheta)=\exp[-\mathrm{i}\vartheta\sigma_\nu/2]$ with $\nu=x,y,z,$ and $\sigma_\nu$ are the Pauli matrices. 
The encoded state then passes through $l$ strongly entangling layers. 
Each layer applies a three-parameter rotation to every qubit, followed by a sequence of controlled-NOT (CNOT) gates. 
For qubit $q$ in layer $\ell$, the rotation is
\begin{equation}
\mathcal R_{q\ell}
=R_z(\alpha_{q\ell3})R_y(\alpha_{q\ell2})R_z(\alpha_{q\ell1}),
\label{eq:hqnnrot}
\end{equation}
with the rotation operators acting from right to left.
The CNOT gates couple the qubits and can generate entanglement. 
With the first qubit as control and the second as target, their two-qubit representation is
\begin{equation}
\begin{aligned}
\operatorname{CNOT}
&=|0\rangle\langle0|\otimes\mathbb I_2
  +|1\rangle\langle1|\otimes\sigma_x\\
&=\begin{pmatrix}
1&0&0&0\\
0&1&0&0\\
0&0&0&1\\
0&0&1&0
\end{pmatrix},
\end{aligned}
\label{eq:hqnncnot}
\end{equation}
where $\mathbb I_2$ is the single-qubit identity.
Thus, the target is flipped only when the control is in $|1\rangle$. 
These gates introduce no trainable parameters.
In summary, the operations defined in Eqs.~(\ref{eq:hqnnrot}) and~(\ref{eq:hqnncnot}) together form a single quantum circuit layer.
We denote the overall unitary transformation of the quantum circuit by $V(\bm{\alpha})$, where $\bm{\alpha}$ is the set of trainable circuit parameters.

After the final layer, the quantum state and measured features are
\begin{equation}
\begin{aligned}
|\psi\rangle&=V(\bm\alpha)U(\bm\theta)|\bm 0\rangle,\\
\mathcal{Q}_q&=\langle\psi|\sigma_z^{(q)}|\psi\rangle,
\qquad q=1,\ldots,n,
\end{aligned}
\label{eq:hqnnmeasurement}
\end{equation}
where $\sigma_z^{(q)}$ acts as the Pauli-$Z$ operator on qubit $q$ and as the identity operator on all other qubits.
Each expectation value lies in $[-1,1]$. 
The resulting vector $\mathbf{Q} = (\mathcal{Q}_1,\ldots,\mathcal{Q}_n)^T$ is passed to an $n\to32\to1$ classical readout,
\begin{equation}
\widetilde r_p
=W_4\operatorname{SiLU}(W_3 \mathbf{Q}+\bm b_3)+b_4,
\label{eq:hqnnreadout}
\end{equation}
where $W_3,W_4$ and $\bm b_3,b_4$ are its trainable weights and biases, and $\widetilde r_p$ is the predicted standardized residual. 

The rotations in Eq.~\eqref{eq:hqnnrot} represent arbitrary single-qubit operations up to a global phase.
Together with CNOT gates, they form a universal gate set: other unitary gates can be synthesized from suitable sequences of these operations~\cite{barenco1995gates}.
However, since the depth of quantum circuits is often constrained in practical implementations, alternative entangling gates, coupling patterns, and circuit architectures warrant further investigation for their impacts on trainability and nuclear mass prediction accuracy.

To guide the choice of qubit number $n$ and circuit depth $l$, we perform a preliminary architecture study using a spherical Skyrme–Hartree–Fock + Bardeen–Cooper–Schrieffer (SHF+BCS) baseline~\cite{reinhard1991shf}.
The relatively large residuals of this baseline provide a useful setting for distinguishing the performance of different circuit configurations.
Based on this study, we adopt $n=l=6$ as a reasonable configuration.
See Appendix~\ref{app:legacyA} for details.
The encoder, circuit, and readout contain 486, 108, and 257 trainable parameters, respectively, for a total of 851.
For comparison, we also design a classical expert, labeled ANN in the subsequent figures.
It uses the same descriptors and residual target and also contains 851 trainable parameters, matching the HQNN count.
In practice, we tested ANN architectures with different depths and widths.
At the current parameter scale, their performances were comparable, and we present only one representative architecture ($8\to85\to1$).

The HQNNs are implemented using PyTorch~\cite{paszke2019pytorch} and PennyLane~\cite{bergholm2018pennylane}, enabling joint optimization of the classical weights and quantum circuit parameters within each expert. 
The trained circuit parameters can be reused on compatible quantum processors after hardware-specific circuit compilation. 
In such a deployment, the quantum circuits execute on quantum hardware, while the encoder, readout, and gating networks run on a classical computer.

\subsection{Gating and ensemble predictions}
\label{sec:gateinputs}

To combine the predictions of sixteen HQNN experts, we adopt a MoE framework in which a trainable gating network assigns nucleus-dependent weights to the expert outputs.
The gating network maps the standardized inputs $\widetilde{\bm{u}}_{\mathrm{MoE}}$ to gating logits $g_j$ for each expert $j$, which are converted into normalized weights,
\begin{equation}
w_j=\frac{\exp(g_j/T)}{\sum_{j'=1}^{16}\exp(g_{j'}/T)},
\label{eq:gateA}
\end{equation}
where $T=2.5$ is the softmax temperature.
The resulting residual prediction is $\widetilde r_{\mathrm{MoE}}=\sum_{j=1}^{16}w_j\widetilde r_j$.
Since the weights are nonnegative and sum to one, the MoE prediction lies within the range of the expert predictions for each nucleus.
We train sixteen gating networks using the same fixed pool of sixteen experts and average the resulting MoE predictions to obtain the final result.

For the gating networks, we initially used the same eight input descriptors as for the HQNN experts. 
Feature-importance analysis highlighted the role of shell-related information, motivating a more explicit shell descriptor (Appendix~\ref{app:featureranking}). 
To further examine Coulomb and surface effects and allow for an additional nonlinear dependence on nuclear size, we introduced four additional features: the shell descriptor $H=d_Zd_N/(d_Z+d_N+\epsilon)$, the Coulomb descriptor $C=Z(Z-1)/(A^{1/3})$, the surface descriptor $A^{2/3}$, and the quadratic mass-number term $A^2$. 
Here \(\epsilon\) is a small positive constant introduced to avoid division by zero.
The resulting twelve-dimensional input vector is
\begin{equation}
\begin{split}
{\bm{u}}_{\mathrm{MoE}}={{\bm{u}}}\oplus \{H,C,A^{2/3},A^2\}.
\end{split}
\label{eq:gatefeaturesA}
\end{equation}
These descriptors supply nuclear-structure information to the gating network. 
The feature scores $I_k$ are defined in Eq.~\eqref{eq:importanceA}; the global feature comparison is reported in Appendix~\ref{app:featureranking}.
The classical gating network has hidden-layer widths of 32, 128, 256, 512, 256, and 64, and produces sixteen output logits. 
With the expert parameters held fixed, each gating network is trained by minimizing the mean-squared error of the combined residual prediction,
\begin{equation}
\mathcal{L}_{\mathrm{MoE}}=\frac{1}{n_{\mathrm{tr}}}\sum_{i=1}^{n_{\mathrm{tr}}}\left[\widetilde r_{\mathrm{MoE}}(Z_i,N_i)-\widetilde r(Z_i,N_i)\right]^2,
\label{eq:moelossA}
\end{equation}
where $n_{\mathrm{tr}}$ is the number of training nuclei.

In this study, we train sixteen models using data for 3193 nuclei with $Z\geq20$, including both experimentally determined masses and extrapolated estimates reported in AME2020~\cite{wang2021ame}.
Including these extrapolated masses leads to a somewhat larger WS4 error relative to the AME2020 reference values than that obtained using experimentally determined masses alone.
The data are divided into training, validation, and test sets in a $6:3:1$ ratio, with the same test set held out for independent evaluation of all models.
The remaining nuclei constitute the development set, which is repartitioned into training and validation subsets across models for cross-validation.
For each training run, we retain the model with the lowest validation root-mean-square error (RMSE).

\section{Nuclear mass predictions}

\begin{figure}[!tb]
\centering
\includegraphics[width=\columnwidth]{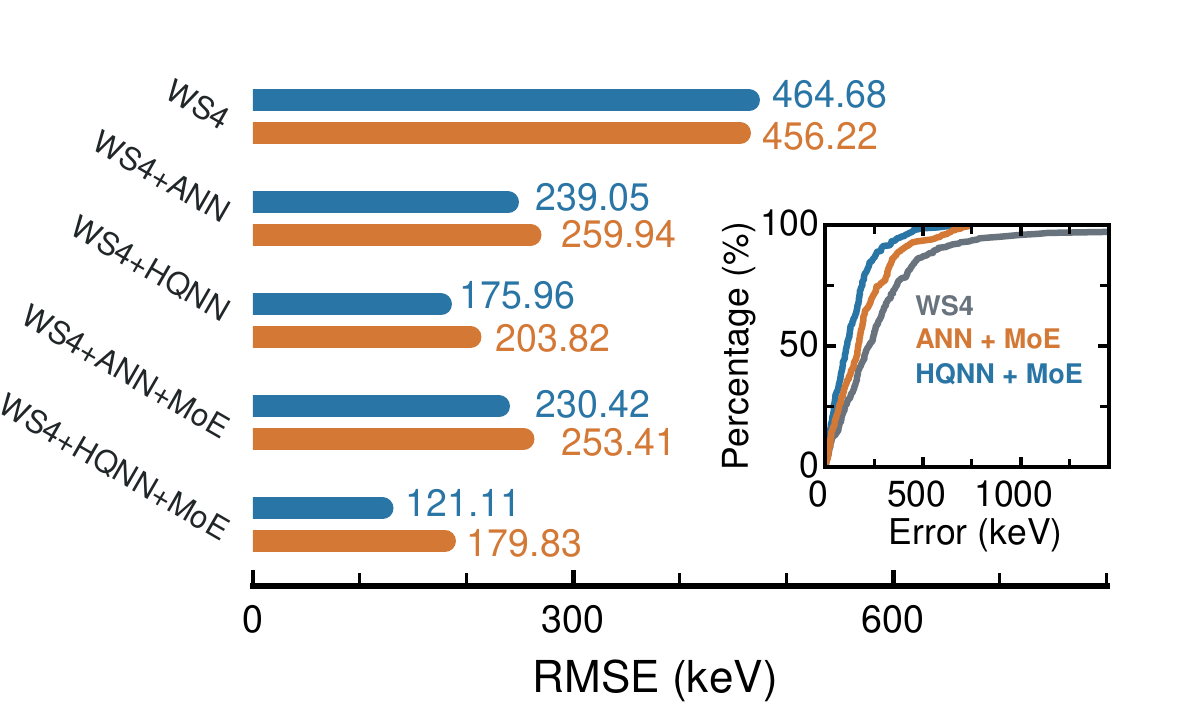}
\caption{RMSEs relative to reference masses for WS4, uniformly averaged sixteen-expert ensembles, and ensembles of sixteen MoE models. 
Blue bars represent the development pool and orange bars represent the fixed test set. 
The inset shows cumulative distributions of absolute test residuals.}
\label{fig:main2}
\end{figure}

We first assess the global accuracy of the predicted nuclear binding energies. 
Figure~\ref{fig:main2} compares the development- and test-set RMSEs for five predictors: WS4, the uniformly averaged ANN and HQNN ensembles, and their respective MoE extensions.

With WS4 providing the underlying binding-energy surface, the expert models learn corrections to its remaining nucleus-dependent deviations from the reference values. 
Uniform averaging reduces the test RMSE from 456.22~keV for WS4 to 203.82~keV for HQNN and 259.94~keV for ANN, respectively. 
These reductions suggest that the baseline residuals contain systematic patterns that can be learned from the nuclear features. 
The HQNN ensemble provides more accurate corrections than the ANN ensemble even before nucleus-dependent MoE weighting is introduced.
Furthermore, the MoE reduces the HQNN test RMSE to 179.83~keV. 
Adaptive weighting thus improves the combined residual correction for the hybrid expert pool. 
The ANN test RMSE also decreases but just slightly, from 259.94 to 253.41~keV after gating. 
The HQNN also outperforms the other parameter-matched ANN architectures tested with different depths and widths.
The comparison therefore identifies two distinct gains for HQNN--MoE: more accurate expert predictions and a further improvement from their adaptive combination. 

The inset of Fig.~\ref{fig:main2} complements this comparison with the cumulative distributions of absolute test errors for WS4, HQNN--MoE, and ANN--MoE. 
At each error threshold, the vertical coordinate gives the percentage of test nuclei with errors below that threshold. 
The HQNN--MoE curve rises more rapidly at small errors, showing that a larger fraction of nuclei is reproduced with small deviations from the reference values. 
Its lower RMSE is therefore accompanied by smaller typical errors, rather than a reduction in a few extreme residuals.
Figure~\ref{fig:main5} in Sec.~\ref{sec:driplines} locates these remaining mass errors and relates the corrected mass surfaces to predicted drip lines.

The gain from adaptive weighting motivates examining which nuclear descriptors guide the gating networks across the nuclear chart.
We use normalized logit-gradient scores to characterize their relative use of these descriptors. 
For a single MoE model, let $\bm{x}_i\equiv\widetilde{\bm{u}}_{\mathrm{MoE},i}$ denote the standardized input vector for nucleus $i$, with $x_{ik}$ its $k$th feature component. 
We first average the absolute gradients of the gating logits $g_j$ over the $M = 3193$ reference nuclei,
\begin{equation}
G_{jk}=\frac{1}{M}\sum_{i=1}^{M}\left|\frac{\partial g_j(\bm{x}_i)}{\partial x_{ik}}\right|.
\end{equation}
For each expert, this gradient profile is normalized over the $K$ input features and then combined with the nucleus-dependent expert weights,
\begin{equation}
I_k(i)=\sum_{j=1}^{16}w_{ji}\frac{G_{jk}}{\sum_{k'=1}^{K}G_{jk'}},
\label{eq:importanceA}
\end{equation}
where $w_{ji}=w_j(\bm{x}_i)$ follows Eq.~\eqref{eq:gateA}, and $K=8$ or $12$ for the two input schemes. 
Averaging over the gating networks and nuclei gives the global ranking.
Because $G_{jk}$ is averaged over the reference nuclei, spatial variation in $I_k(i)$ enters through the nucleus-dependent weights $w_{ji}$.

The global ranking places $H$, $d_Z$, and $d_N$ highest in the twelve-feature model, motivating the spatial analysis of these three scores.
The numerical rankings and the eight- versus twelve-feature comparison are given in Fig.~\ref{fig:a2} and Appendix~\ref{app:featureranking}.

\begin{figure}[!tbp]
\centering
\includegraphics[width=\columnwidth,height=0.69\textheight,keepaspectratio]{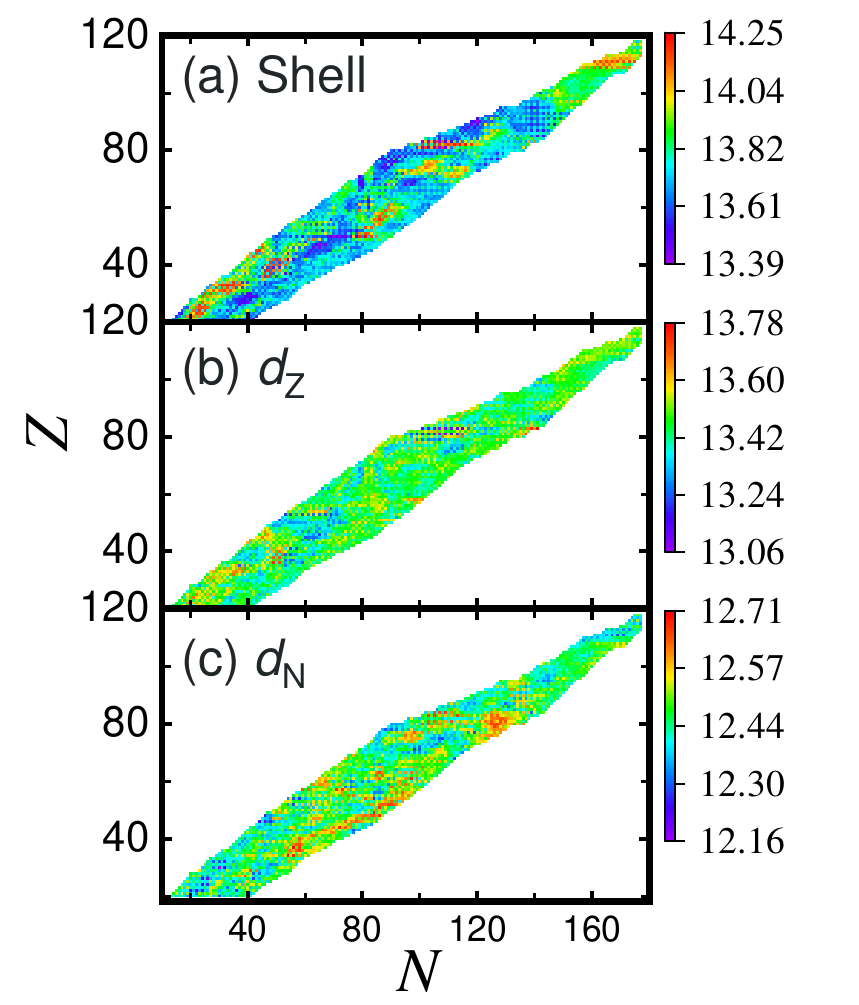}
\caption{Spatial distributions of the three highest-ranked feature scores for the twelve-input HQNN--MoE gating networks: (a) $H$ (labeled Shell), (b) $d_Z$, and (c) $d_N$.
Colors show the scores averaged over the gating networks, expressed as percentages.}
\label{fig:main3}
\end{figure}

Figure~\ref{fig:main3} shows how the three highest-ranked feature scores vary across the nuclear chart.
Panel (a) shows the score associated with the combined shell descriptor $H$, while panels (b) and (c) show those associated with $d_Z$ and $d_N$, respectively.
Panel (a) exhibits faint bands near magic proton or neutron numbers, whereas the distributions in panels (b) and (c) appear smoother.
Odd--even modulation is also visible in the scores across the nuclear chart.
These spatial patterns arise from nucleus-dependent expert weighting learned by the gating networks trained on the WS4 residuals.
The prominence of shell-related descriptors and the odd--even modulation motivate a closer examination of shell- and pairing-dependent mass variations.
This finding provides important guidance for the further development of the WS series of models and suggests that greater attention should be paid to shell effects.

We therefore turn from descriptor use to the consistency of the predicted energies across neighboring nuclei.
A lower global mass RMSE does not by itself ensure accurate local mass differences.
Errors in neighboring masses can cancel or reinforce when separation energies are formed.
We compare separation energies and higher-order mass filters for the isotope chains of magic-proton Ca, Ni, Sn, and Pb.
The one- and two-neutron separation energies are
\begin{align}
S_n(Z,N)&=B(Z,N)-B(Z,N-1),\label{eq:one_neutronA}\\
S_{2n}(Z,N)&=B(Z,N)-B(Z,N-2).
\label{eq:separationA}
\end{align}
The quantities $S_n$ and $S_{2n}$ measure the energies required to remove one and two neutrons, respectively, and characterize neutron binding along an isotope chain.
The odd--even staggering of $S_n$ reflects neutron pairing.
Since $S_{2n}$ compares nuclei with the same neutron-number parity, it reduces this staggering and provides a smoother view of the evolution of neutron binding.
Kinks or abrupt decreases in these trends can reveal shell closures and changes in nuclear deformation~\cite{buskirk2024shells}.

At fixed $Z$, the change in $S_{2n}$ over a two-neutron step defines the empirical two-neutron shell-gap indicator,
\begin{equation}
\delta_{2n}(Z,N)=S_{2n}(Z,N)-S_{2n}(Z,N+2).
\end{equation}
At a shell closure, the occupation of more weakly bound orbitals above the gap leads to a pronounced decrease in $S_{2n}$ and a corresponding enhancement of $\delta_{2n}$.
This indicator therefore probes shell structure through local changes in the binding-energy surface, while also retaining contributions from changes in pairing and deformation correlations~\cite{buskirk2024shells,bender2006quadrupole}.

A complementary probe of neutron single-particle structure is provided by
\begin{equation}
\Delta e_n(Z,N)=S_n(Z,N)-S_n(Z,N+2).
\end{equation}
For even $N$, this difference compares one-neutron separation energies of the same neutron-number parity, reducing their common pairing contribution.
When the mean field and pairing vary smoothly, $\Delta e_n$ approximates the effective spacing between the last occupied and next available neutron orbitals.
Its isotope dependence thus provides information on shell and subshell gaps near the Fermi surface~\cite{satula1998staggering,buskirk2024shells}.

To resolve more localized variations in the shell-gap pattern, we further consider
\begin{align}
H_{4n}(Z,N)&=\delta_{2n}(Z,N)\nonumber\\
&\quad-\frac{\delta_{2n}(Z,N-2)+\delta_{2n}(Z,N+2)}{2}.
\label{eq:massfilters}
\end{align}
This fourth-order mass filter measures the excess of the central $\delta_{2n}$ value over the average of its two neighbors.
It suppresses a smooth local background and emphasizes localized enhancements in the shell-gap structure.
Together, these observables connect the evolution of neutron binding with shell structure and its interplay with pairing and deformation.

\begin{figure*}[tbh]
\centering
\includegraphics[width=0.88\textwidth,height=0.79\textheight,keepaspectratio]{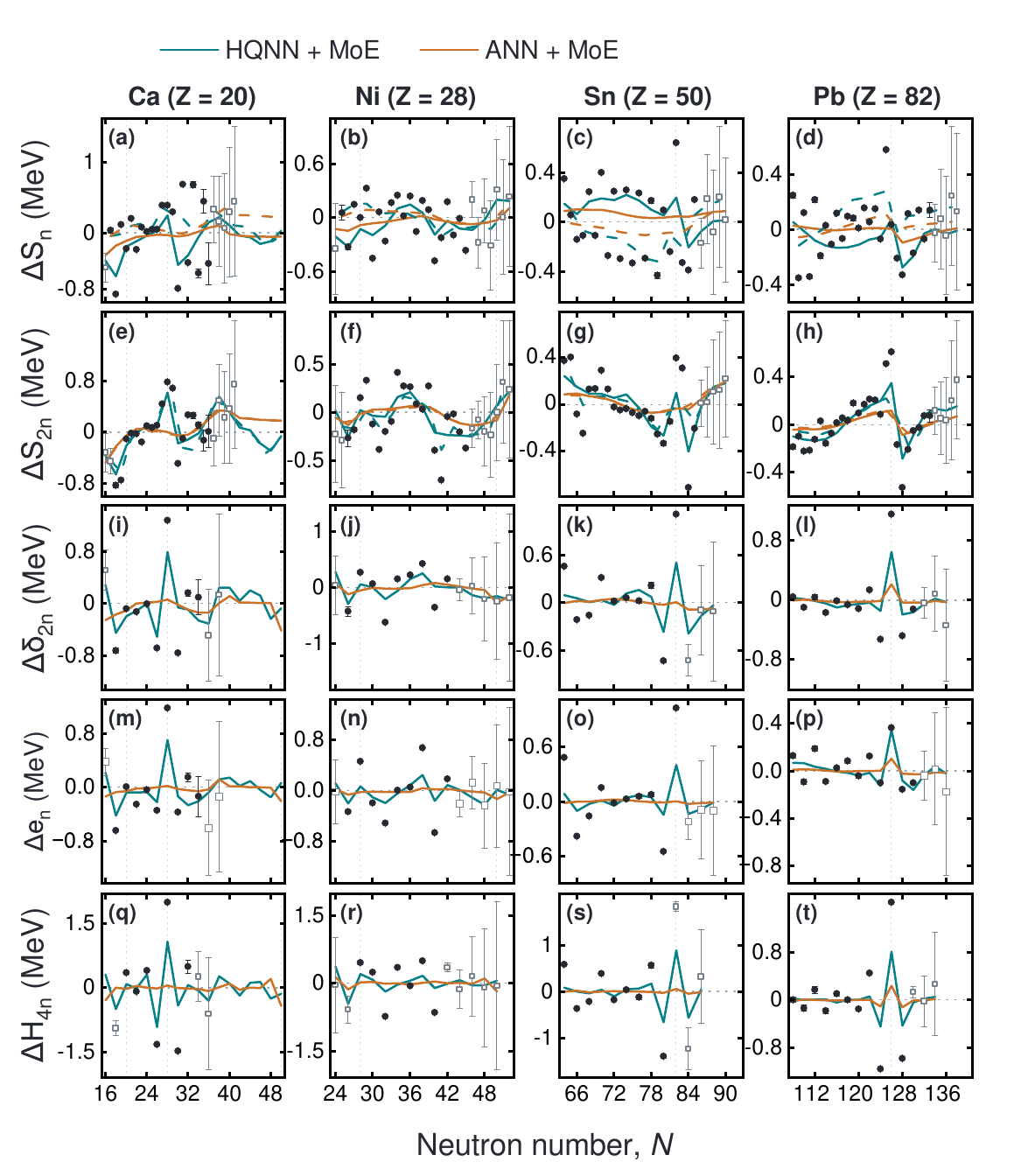}
\caption{Corrections relative to WS4 along the Ca, Ni, Sn, and Pb isotope chains, arranged from left to right.
From top to bottom, the rows show corrections to $S_n$, $S_{2n}$, $\delta_{2n}$, $\Delta e_n$, and $H_{4n}$.
WS4 is subtracted from both the model predictions and the AME2020 reference values.
Teal and orange curves denote HQNN--MoE and ANN--MoE, respectively.
In the separation-energy panels, solid and dashed curves correspond to even and odd $N$, respectively; the higher-order filters are evaluated at even $N$.
Black filled circles denote reference corrections constructed entirely from unflagged AME2020 masses, while open gray squares indicate combinations containing at least one estimated mass.
Error bars propagate the quoted mass uncertainties.}
\label{fig:main4}
\end{figure*}

To examine how residual learning modifies these structural signatures, Fig.~\ref{fig:main4} compares the corrections $\Delta Y=Y-Y_{\mathrm{WS4}}$ obtained from the model predictions and AME2020 reference values for each observable $Y$.
Positive and negative values indicate increases and decreases relative to WS4, respectively.
The comparison tests whether the learned residuals recover the local binding-energy variations required by AME2020.
In this representation, the zero line denotes the uncorrected WS4 prediction.
Neutron-emission thresholds are determined from the zeros of the full separation energies $S_n$ and $S_{2n}$.

The separation-energy panels in Fig.~\ref{fig:main4}(a)--(h) show how the learned corrections modify neutron binding between neighboring isotopes.
For Ca at $N=28$, AME2020 requires an $S_{2n}$ correction of $+0.785$~MeV.
HQNN--MoE predicts $+0.615$~MeV, whereas ANN--MoE only predicts $+0.004$~MeV.
The hybrid model therefore reproduces much of the required increase in $S_{2n}$, while the classical model leaves the WS4 prediction almost unchanged.
This example shows that HQNN--MoE captures the local variation of the binding-energy correction relevant to neutron removal.

The higher-order filters in panels (i)--(t) examine how these corrections modify the shell-gap systematics.
The effect is particularly evident in $\delta_{2n}$ near the Sn and Pb shell closures.
At $N=82$ in Sn, the reference correction is $+1.112$~MeV, compared with $+0.509$~MeV for HQNN--MoE and $+0.005$~MeV for ANN--MoE.
At $N=126$ in Pb, the corresponding values are $+1.145$, $+0.637$, and $+0.207$~MeV, respectively.
The positive reference corrections indicate that WS4 underestimates the empirical two-neutron shell gaps at $^{132}\mathrm{Sn}$ and $^{208}\mathrm{Pb}$, and HQNN--MoE recovers a substantial part of the missing enhancement.
The $\Delta e_n$ and $H_{4n}$ panels provide complementary information on how the corrections affect the effective neutron-level spacings and the local prominence of the shell-gap pattern.
Despite these improvements, HQNN--MoE still shows discrepancies in some local mass differences, indicating room for further refinement in capturing nuclear-structure effects.

Appendix~\ref{app:correctionagreement} extends this analysis to all eligible nuclei, including those in the development set, by comparing the model and reference corrections for each observable.
The comparison shows that HQNN--MoE captures higher-order variations in the mass residuals, whereas ANN--MoE produces corrections concentrated near zero and largely misses these local variations.
This contrast highlights the ability of hybrid quantum--classical learning to resolve finer structure in the residual mass surface, extending the improvement from overall binding-energy accuracy to mass differences sensitive to shell effects.

Higher-order mass differences suppress smooth contributions and emphasize localized changes associated with shell structure and variations in pairing and deformation correlations~\cite{hove2013massrelations,buskirk2024shells}.
In particular, $\delta_{2n}$ probes the local curvature of the binding-energy surface, while $H_{4n}$ involves a fourth-order finite difference, making these observables sensitive to fine variations that a smooth residual correction can miss.
Spectral bias toward low-frequency components has been demonstrated in feedforward networks, particularly for ReLU activations~\cite{rahaman2019spectral}.
(Since SiLU is a smooth variant of ReLU, the spectral bias toward low-frequency components observed in ReLU networks may also be relevant here.)

In the HQNN, angle encoding provides trigonometric features, and entangling operations allow products of these features to contribute to the measured quantum outputs~\cite{mitarai2018qcl,schuld2021encoding}.
Combined with the trainable classical encoder and readout, these coupled features provide a plausible mechanism for retaining local residual variations that are attenuated in the smoother ANN corrections.
The improved higher-order mass differences are consistent with this representation capturing more of the local structure needed to correct the WS4 mass surface.

\section{Nuclear drip lines and particle-emission thresholds}
\label{sec:driplines}

\begin{figure*}[t]
\centering
\begin{minipage}[c]{0.75\textwidth}
\includegraphics[width=\linewidth]{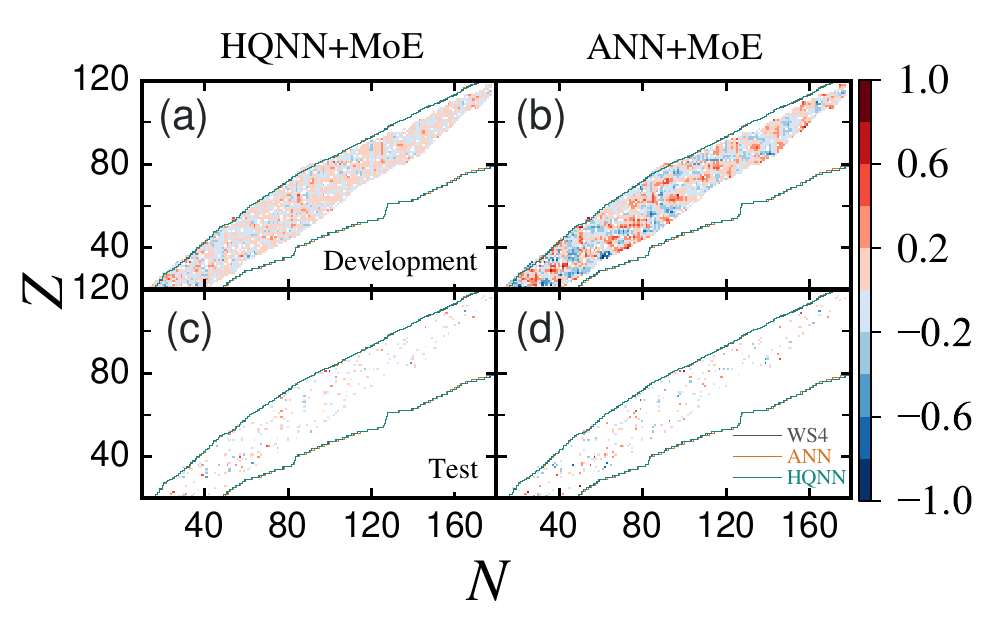}
\end{minipage}\hfill
\begin{minipage}[c]{0.22\textwidth}
\caption{Mass errors and predicted two-nucleon drip lines across the nuclear chart.
Colors show $B_{\mathrm{AME}}-B_{\mathrm{model}}$ in MeV for HQNN--MoE (left) and ANN--MoE (right), with development nuclei above and test nuclei below.
Positive values indicate underbinding.
Gray, orange, and teal curves show the $S_{2n}=0$ and $S_{2p}=0$ boundaries for WS4, ANN--MoE, and HQNN--MoE, respectively.}
\label{fig:main5}
\end{minipage}
\end{figure*}

The preceding mass-difference analysis tests local structure within the reference region.
We now examine how the corrected mass surfaces extend toward the limits of nuclear binding.
Figure~\ref{fig:main5} connects the spatial distribution of mass errors to the predicted two-neutron and two-proton drip lines.
The error maps provide the spatial counterpart of the global comparison in Fig.~\ref{fig:main2}.
HQNN--MoE shows fewer pronounced positive and negative deviations than ANN--MoE across the sampled region, consistent with its lower mass RMSE.

The overlaid curves address a different question: where the predicted separation energies vanish beyond this reference region.
The neutron-rich boundary is obtained from $S_{2n}(Z,N)$ in Eq.~\eqref{eq:separationA}, evaluated at even $N$ along each isotopic chain.
The proton-rich boundary uses $S_{2p}(Z,N)=B(Z,N)-B(Z-2,N)$ at even $Z$ along each isotonic chain.
For each model, we locate the outermost positive-to-nonpositive crossing.
The three models yield similar overall boundaries over the plotted range, although their local zero crossings differ.
Thus, the learned corrections modify selected threshold locations while largely retaining the global extent of the WS4 mass surface.

Near a separation-energy zero crossing, even a small correction to the nuclear mass surface can open or close a neutron-emission channel. 
A positive separation energy indicates that the corresponding removal process requires energy, whereas a negative value makes it energetically allowed. 
To place our predictions in a broader context, we apply the same separation-energy criterion to the FRDM(2012), HFB-24, UNEDF2, BCPM, and Gogny-D1M mass tables~\cite{moller2016frdm,goriely2013hfb24,kortelainen2014unedf2,baldo2013bcpm,goriely2009d1m}. 
The resulting agreement varies considerably across isotopic chains.

For Ni and Zr, both HQNN–MoE and ANN–MoE preserve the WS4 endpoints at $^{86}\mathrm{Ni}$ and $^{122}\mathrm{Zr}$, respectively. 
These predictions coincide with Gogny-D1M and lie at the least neutron-rich limits of the ranges spanned by the five reference models, from $^{86}\mathrm{Ni}$ to $^{96}\mathrm{Ni}$ and from $^{122}\mathrm{Zr}$ to $^{134}\mathrm{Zr}$.
More pronounced differences emerge for Sn, where both corrected models predict an endpoint at $^{158}\mathrm{Sn}$, compared with $^{174}\mathrm{Sn}$ in UNEDF2 and $^{176}\mathrm{Sn}$ in HFB-24, BCPM, and Gogny-D1M. 
For Pb, whose boundary lies beyond the neutron-number range displayed in Fig. 5, HQNN–MoE retains the $^{266}\mathrm{Pb}$ endpoint, in agreement with all five reference models, whereas ANN–MoE shifts it to $^{264}\mathrm{Pb}$.

Calcium provides a particularly informative near-threshold case, where the sensitivity to residual corrections becomes directly apparent. 
Published Ca predictions illustrate the sensitivity of this region to the treatment of nuclear correlations.
Bayesian averaging of corrected global mass models assigns a substantial probability for two-neutron binding in $\Ca{68}$~\cite{neufcourt2019}.
Gamow shell-model calculations with continuum coupling place the drip line at $\Ca{70}$~\cite{li2020gsm}.
A shell-model study with chiral forces also supports two-neutron binding up to $\Ca{70}$~\cite{coraggio2020}.
An exact pairing calculation with continuum correlations and an isospin-dependent interaction instead locates the two-neutron drip line at $\Ca{66}$~\cite{dassie2020}.
A bound $\Ca{66}$ halo scenario has also been investigated with a relativistic continuum model and Lipkin--Nogami pairing~\cite{zhou2024}.

\begin{figure}[tb]
\centering
\includegraphics[width=\columnwidth]{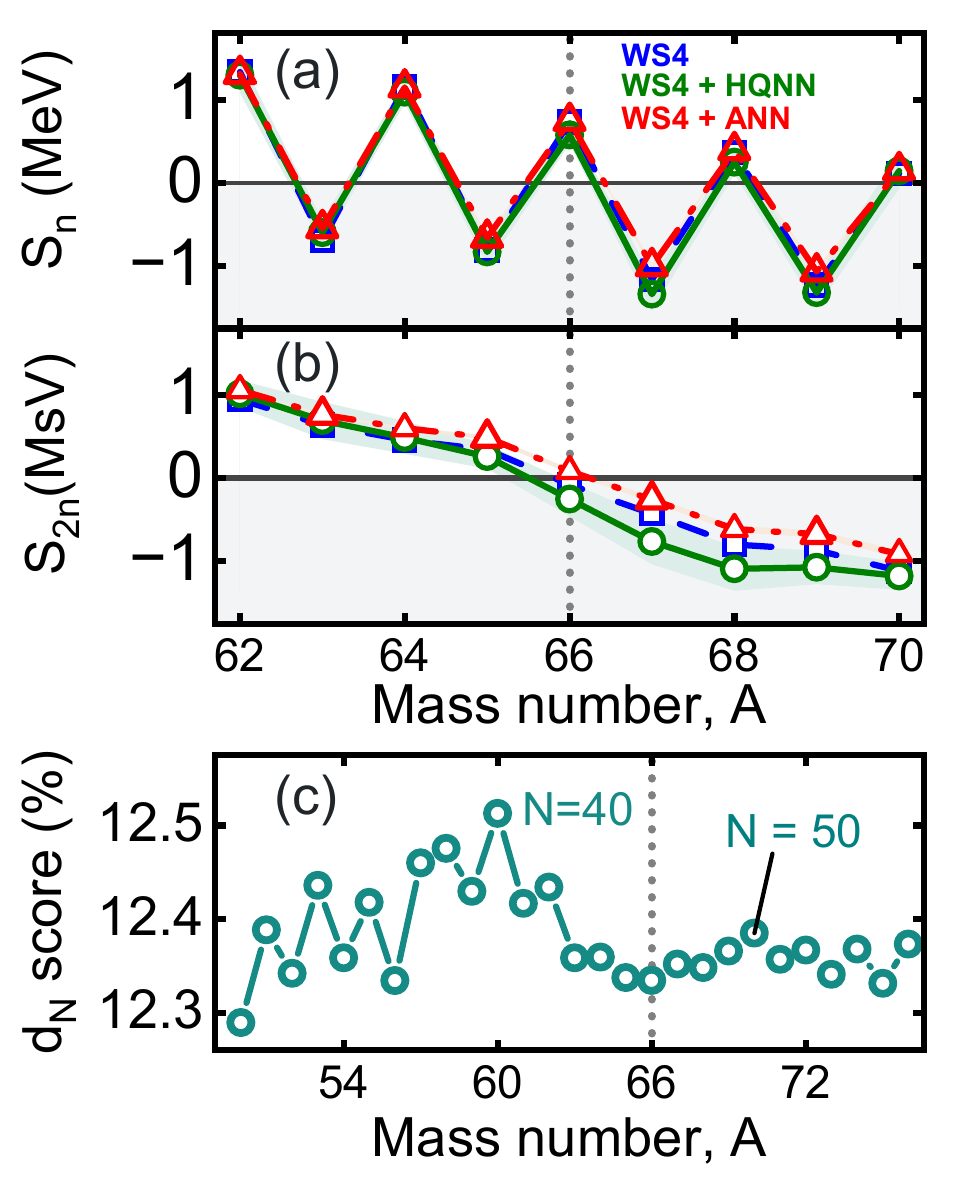}
\caption{Calcium separation energies and the neutron shell-distance feature score in HQNN--MoE. 
(a) One-neutron and (b) two-neutron separation energies for $62\leq A\leq70$. 
Blue squares, green circles, and red triangles denote WS4, HQNN--MoE, and ANN--MoE, respectively.  
(c) The score $I_{d_N}$, averaged over the HQNN--MoE gating networks, for $50\leq A\leq76$. 
Dotted lines mark $\Ca{66}$.
Shaded bands show the standard deviation across the predictions of the  MoE models.}
\label{fig:main6}
\end{figure}

\begin{samepage}
Figure~\ref{fig:main6} resolves the WS4, HQNN--MoE, and ANN--MoE neutron thresholds along the Ca chain, with $\Ca{66}$ providing a direct comparison.
Figure~\ref{fig:main6}(a) shows the one-neutron separation energies for all isotopes from $A=62$ to 70.
All three curves alternate between higher values for even-$A$ nuclei and lower values for their odd-$A$ neighbors.
At $\Ca{66}$, the predicted $S_n$ values are 0.733~MeV for WS4, 0.577~MeV for HQNN--MoE, and 0.680~MeV for ANN--MoE.
All three values remain above the zero-energy line.
\par
\end{samepage}

The two-neutron separation energies in panel (b) decrease more smoothly along the chain.
All three models give positive $S_{2n}$ at $\Ca{64}$ and negative $S_{2n}$ at $\Ca{68}$.
Their predictions for the intervening even nucleus $\Ca{66}$ lie close to zero but have different signs.
WS4 gives $-0.077$~MeV and HQNN--MoE gives $-0.252$~MeV, whereas ANN--MoE gives $+0.115$~MeV.
ANN--MoE therefore keeps both neutron-removal channels closed at $\Ca{66}$, whereas WS4 and HQNN--MoE allow two-neutron emission while keeping one-neutron emission closed.
For HQNN--MoE, the standard deviations across the sixteen gating-network realizations at $\Ca{66}$ are $0.106$~MeV for $S_n$ and $0.205$~MeV for $S_{2n}$.
The corresponding one-standard-deviation intervals retain $S_n>0$ and $S_{2n}<0$, although the upper limit of the $S_{2n}$ interval lies close to zero.
The $S_{2n}$ uncertainty band is broader around $\Ca{67}$ and $\Ca{68}$, with standard deviations of approximately $0.28$ and $0.26$~MeV, respectively.

Panel (c) shows the mean neutron shell-distance score $I_{d_N}$ over the wider range $50\leq A\leq76$.
Since $Z=20$ is a magic number, the input descriptors $H$ and $d_Z$ are zero throughout the calcium chain.
We therefore focus on the isotope dependence of the neutron shell-distance score.
The curve reaches a local maximum at $\Ca{60}$ ($N=40$), followed by a shallow local minimum at $\Ca{66}$ ($N=46$) and a modest rise toward $N=50$.
The possible subshell closure at $N=40$ has been extensively investigated~\cite{li2020gsm,chen2023}.
The ANN--MoE score in Fig.~\ref{fig:a4} is nearly constant between $N=40$ and $N=50$, without a comparable peak at $N=40$ or dip at $N=46$, and decreases beyond $N=50$.
By Eq.~\eqref{eq:importanceA}, the contrast reflects different isotope-dependent mixtures of the normalized gradient profiles of the gating logits.
Possible reasons for the ANN plateau are discussed in Appendix~\ref{app:dnncaA}.

\section{Conclusion}
We have developed an HQNN--MoE framework that combines the physical baseline of WS4 with adaptive residual corrections from hybrid quantum--classical experts.
Within each expert, parameterized rotations and entangling gates construct nonlinear representations of the nuclear inputs and their correlations.
This circuit structure provides a compact and flexible basis for learning nuclear-mass residuals.
On the fixed 257-nucleus AME2020 test set, which includes extrapolated estimates, the framework reduces the mass RMSE from 456.22 to 179.83~keV.
Its improvement over the parameter-matched ANN--MoE supports the effectiveness of this hybrid representation, while the further gain over uniform expert averaging demonstrates the benefit of nucleus-dependent expert weighting.
Together, these results support hybrid quantum circuits as useful components for refining nuclear mass predictions within an established physical framework.

The Ca, Ni, Sn, and Pb comparisons test whether improved mass predictions also recover local separation-energy and shell-gap variations.
HQNN--MoE captures part of the additional gap indicated by AME2020 at the Sn and Pb shell closures, but misses some other local corrections.
These results distinguish a lower average mass error from uniform accuracy in neighboring-mass differences.

The MoE framework also provides a way to examine the structural information used in residual learning.
The high scores of shell-related descriptors and the odd--even modulation of these scores suggest that shell- and pairing-dependent systematics remain important targets for improving WS4.
Along the calcium chain, the HQNN neutron shell-distance score varies nonmonotonically, with a local maximum near $N=40$ and a shallow minimum at $N=46$, whereas the ANN score remains nearly constant between \(N=40\) and \(N=50\).
This contrast suggests that the hybrid expert pool supports more differentiated use of shell-related information as the neutron configuration changes.
The score patterns identify aspects of the model response for further investigation, rather than establishing a shell-structure mechanism.
Together with the mass-difference diagnostics, they can guide tests of shell evolution and odd--even contributions in future mass models, including WS5.
Across the nuclear chart, the corrected two-nucleon boundaries remain close to WS4, while local shifts expose the sensitivity of threshold predictions to residual learning.
The calcium chain illustrates how these shifts can change individual neutron-emission channels.
For $\Ca{66}$, HQNN--MoE predicts energetically allowed two-neutron emission while one-neutron emission remains closed.

This work provides data-driven guidance for refining nuclear mass models and identifying experimental targets near the neutron drip line.
It also supports further applications of hybrid quantum--classical learning in nuclear physics.

\begin{acknowledgments}
This work was supported in part by the National Natural Science Foundation of China (Grant Nos. 12505145 and 12375126), 
the Fundamental Research Funds for the Central Universities (Grant No. SWU-KR25034), 
and the JSPS Grant-in-Aid (Grant Nos. 26K07063 and 26K01431).
\end{acknowledgments}

\appendix
\section{SHF+BCS architecture study}
\label{app:legacyA}

\begin{figure}[tb]
\centering
\includegraphics[width=\columnwidth]{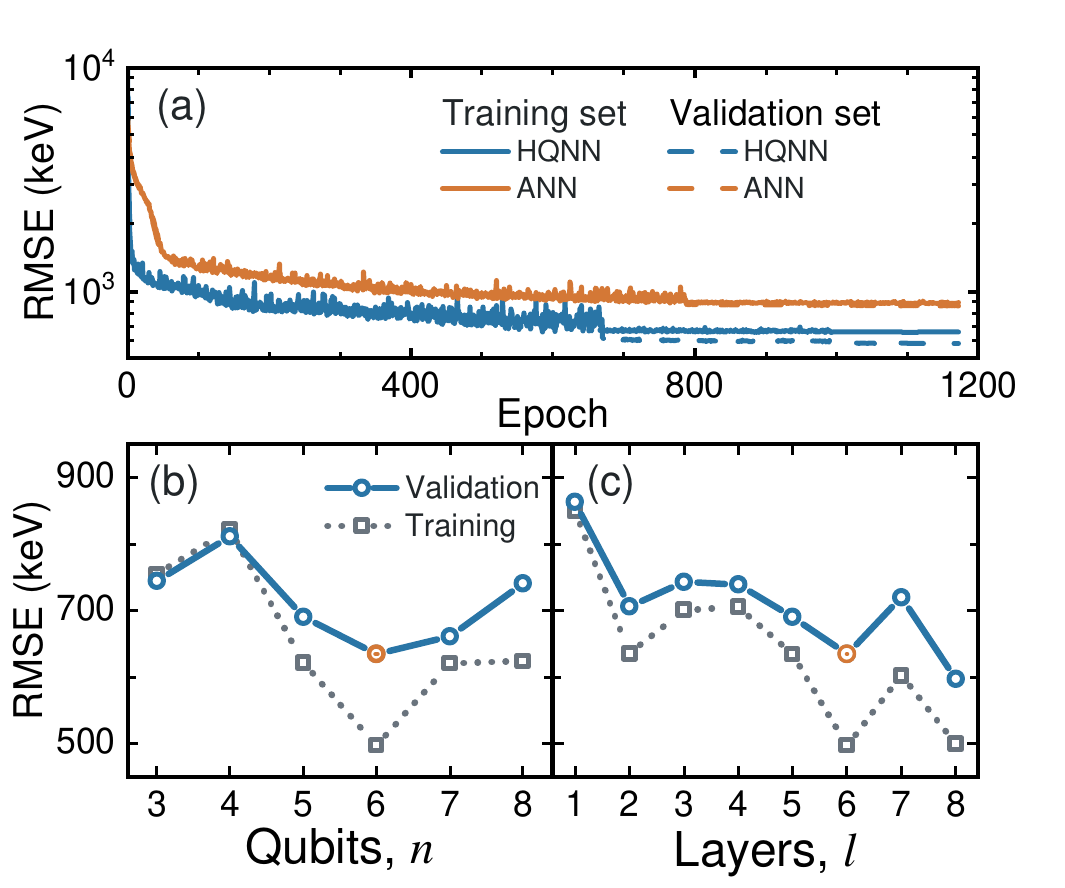}
\caption{Architecture study using an SHF+BCS baseline. (a) Training and validation RMSEs for single HQNN and ANN experts. 
(b) Qubit-number scan with six circuit layers. 
(c) Circuit-depth scan with six qubits. 
}
\label{fig:a1}
\end{figure}

Figure~\ref{fig:a1} presents the single-expert benchmarks used to guide the choice of HQNN architecture.
These calculations use spherical SHF+BCS predictions from Reinhard's program as the baseline~\cite{reinhard1991shf}.
Panel (a) compares the optimization histories of an HQNN and an ANN, with the HQNN achieving a much lower validation RMSE in the comparison shown.

Panels (b) and (c) examine the effects of circuit width and depth separately.
At a fixed depth of six layers, the six-qubit circuit achieves the lowest validation RMSE among the tested widths.
For six qubits, increasing the depth from six to eight layers further reduces the validation error.
We adopt the six-qubit, six-layer architecture as a practical balance between predictive performance and circuit complexity.
This configuration contains 108 trainable rotation angles, giving a total of 851 parameters per expert when the classical encoder and readout are included.

\section{Global feature scores and comparison of gating inputs}
\label{app:featureranking}

\begin{figure}[tb]
\centering
\includegraphics[width=0.87\columnwidth,height=0.64\textheight,keepaspectratio]{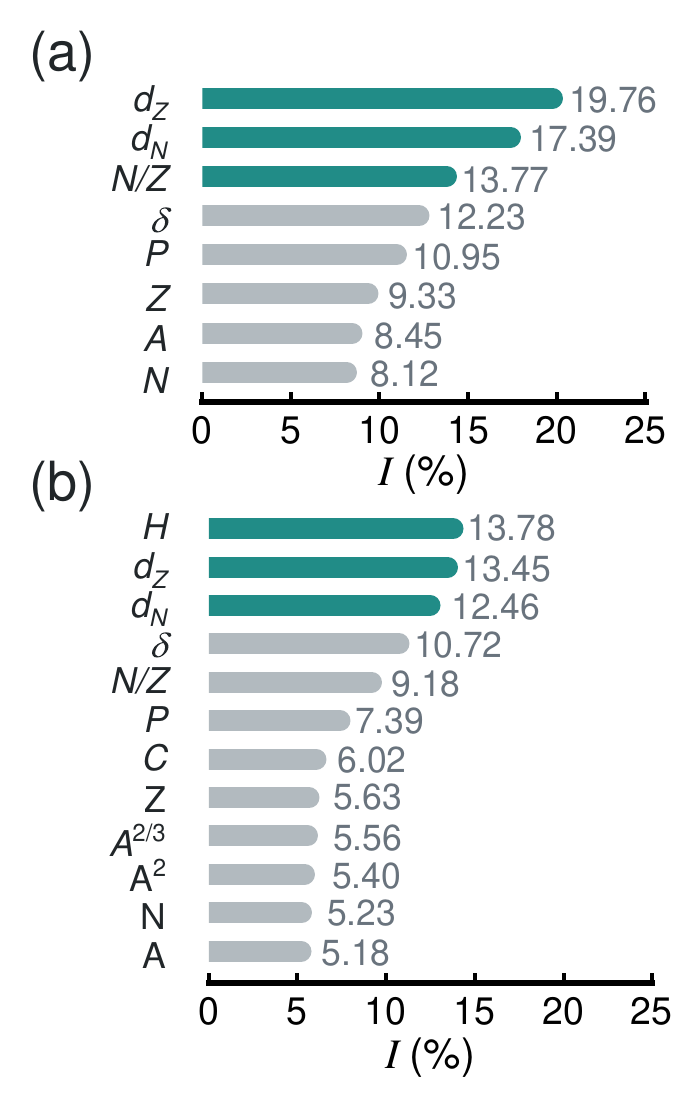}
\caption{Global feature scores for the matched HQNN--MoE ensembles with (a) eight and (b) twelve input features for the gating networks. }
\label{fig:a2}
\end{figure}

This appendix compares the two input-feature sets for the gating networks defined in Sec.~\ref{sec:gateinputs}, using the score in Eq.~\eqref{eq:importanceA}.
For the gating networks with eight input features, Fig.~\ref{fig:a2}(a) ranks $d_Z$ and $d_N$ highest, at 19.76\% and 17.39\%, followed by $N/Z$ at 13.77\%.
This ranking motivates the explicit combined shell descriptor $H$ in the expanded input set.
With the four additional descriptors in Eq.~\eqref{eq:gatefeaturesA}, $H$ ranks first at 13.78\%, followed by $d_Z$ at 13.45\% and $d_N$ at 12.46\%.
Shell-related descriptors therefore remain prominent in both representations.
The twelve-feature ranking supports the choice of the three spatial maps examined in Fig.~\ref{fig:main3}.

The expanded representation changes the ranking more than the predictive accuracy.
The development and test RMSEs decrease from 125.37 and 181.18~keV for eight inputs to 121.11 and 179.83~keV for twelve inputs, respectively.
Because the scores are normalized within each input set, their percentages describe relative descriptor use in that representation.

\section{Global comparison of corrections relative to WS4}
\label{app:correctionagreement}

\begin{figure*}[tbh]
\centering
\includegraphics[width=0.88\textwidth,height=0.79\textheight,keepaspectratio]{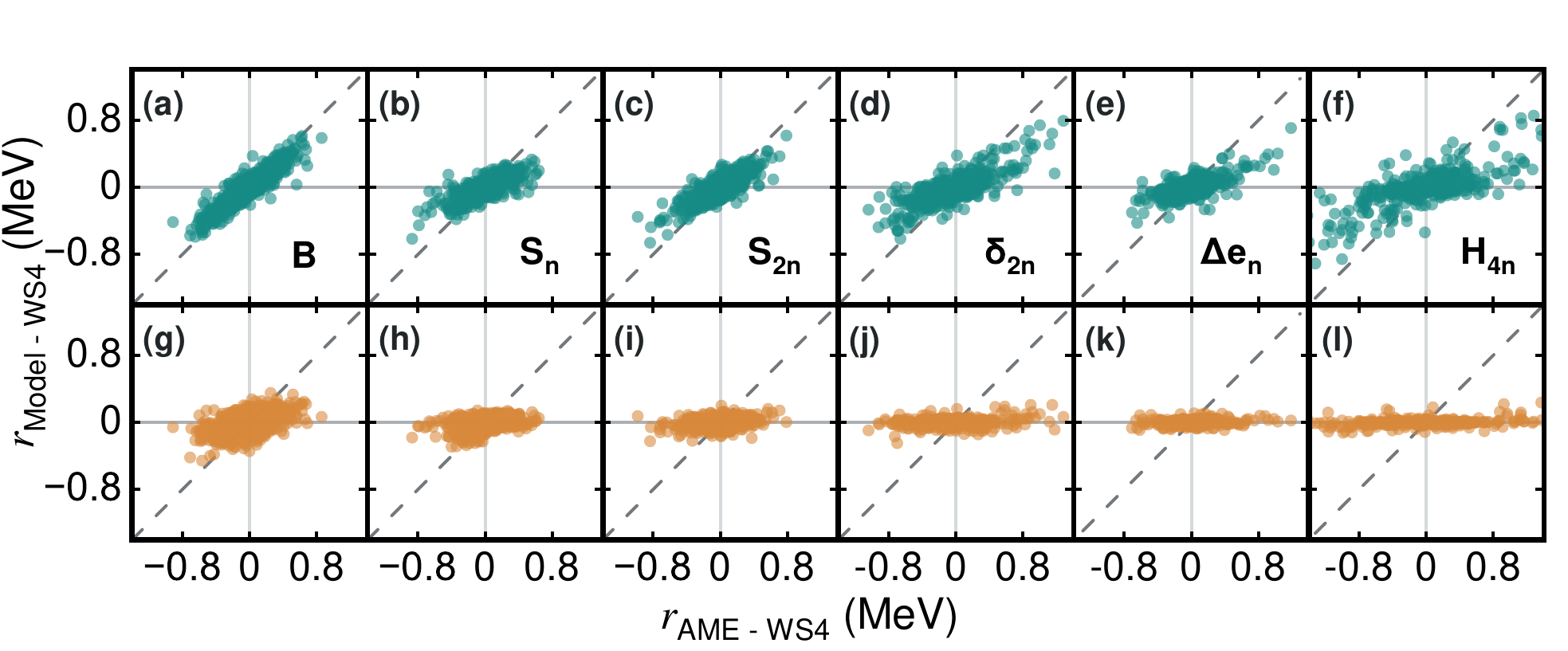}
\caption{Model corrections relative to WS4 versus the corresponding AME2020 corrections.
The columns show the total binding energy $B$, $S_n$, $S_{2n}$, $\delta_{2n}$, $\Delta e_n$, and $H_{4n}$; the top and bottom rows show HQNN--MoE and ANN--MoE, respectively.
For each observable $Y$, the horizontal coordinate is $Y_{\mathrm{AME}}-Y_{\mathrm{WS4}}$ and the vertical coordinate is $Y_{\mathrm{Model}}-Y_{\mathrm{WS4}}$, both in MeV.
Dashed diagonals indicate exact agreement.
The sample includes development and test nuclei.}
\label{fig:a3}
\end{figure*}

Figure~\ref{fig:a3} compares the learned corrections with those required by the reference mass surface.
The binding-energy panels use the total binding energy, and the remaining columns use the definitions in Eqs.~\eqref{eq:one_neutronA}--\eqref{eq:massfilters}.
There are 593 eligible binding-energy points, followed by 546, 545, 499, 494, and 416 complete mass combinations for $S_n$, $S_{2n}$, $\delta_{2n}$, $\Delta e_n$, and $H_{4n}$, respectively.

The HQNN--MoE corrections span a broader range and follow the required corrections more closely in the binding-energy and separation-energy panels.
The ANN--MoE points are more concentrated near zero, particularly for the higher-order mass differences.
On this reference-inclusive sample, HQNN--MoE and ANN--MoE have $\delta_{2n}$ RMSEs of 230.69 and 335.82~keV, respectively.
Their $H_{4n}$ RMSEs are 371.81 and 511.74~keV.
The lower HQNN--MoE errors support improved agreement on these samples, while the scatter around the diagonals and the local counterexamples in Fig.~\ref{fig:main4} show remaining discrepancies.
Subtracting the same WS4 baseline from model and reference leaves their difference unchanged.
This figure is a descriptive comparison that includes development nuclei, and neighboring points can share constituent masses; it therefore supplements the fixed-test-set comparison in Fig.~\ref{fig:main2}.

\section{ANN training protocol and calcium-chain feature scores}
\label{app:dnncaA}

\begin{figure}[tbh]
    \centering
    \includegraphics[width=\columnwidth]{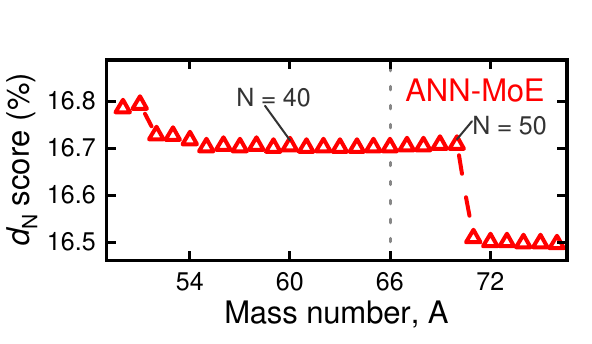}
    \caption{Mean neutron shell-distance score $I_{d_N}$ of the ANN--MoE gating networks along the calcium chain for $50\leq A\leq76$. 
The dotted line marks $\Ca{66}$; labels identify $N=40$ and $N=50$.}
    \label{fig:a4}
\end{figure}

The classical comparison uses sixteen ANN experts, each with a SiLU hidden layer and 851 trainable parameters.
Figure~\ref{fig:a4} applies the same score definition as Fig.~\ref{fig:main6}(c).
The nearly flat ANN score may reflect how the classical experts represent nuclear-structure information in the WS4 residuals.
First, the fixed proton number $Z=20$ gives the calcium isotopes a common proton-shell background, with both $d_Z$ and $H$ remaining zero.
This reduces the variation of shell-related inputs to the neutron sector and may favor similar expert combinations across neighboring isotopes.
Second, neutron shell evolution in this region involves coupled changes in orbital occupations and pairing correlations~\cite{chen2023}.
The ANN experts may absorb these effects into a shared residual trend, maintaining a nearly constant relative emphasis on the neutron shell-distance descriptor.
This interpretation is consistent with the strongly concentrated and stable expert weights observed over the plateau.
The contrasting HQNN response suggests that the two expert pools organize these coupled structural contributions differently, with the hybrid pool allowing more localized changes in their relative weighting.

\bibliography{references}

@article{wang2021ame,
  author = {Wang, Meng and Huang, W. J. and Kondev, F. G. and Audi, G. and Naimi, S.},
  title = {The {AME} 2020 atomic mass evaluation ({II}). Tables, graphs and references},
  journal = {Chinese Physics C},
  volume = {45},
  pages = {030003},
  year = {2021},
  doi = {10.1088/1674-1137/abddaf}
}

@article{wang2014ws4,
  author = {Wang, Ning and Liu, Min and Wu, Xizhen and Meng, Jie},
  title = {Surface diffuseness correction in global mass formula},
  journal = {Physics Letters B},
  volume = {734},
  pages = {215--219},
  year = {2014},
  doi = {10.1016/j.physletb.2014.05.049}
}

@article{niu2022nuclear,
  author = {Niu, Z. M. and Liang, H. Z.},
  title = {{Nuclear mass predictions with machine learning reaching the accuracy required by r-process studies}},
  journal = {Physical Review C},
  volume = {106},
  pages = {L021303},
  year = {2022},
  doi = {10.1103/physrevc.106.l021303}
}

@article{jacobs1991moe,
  author = {Jacobs, Robert A. and Jordan, Michael I. and Nowlan, Steven J. and Hinton, Geoffrey E.},
  title = {Adaptive mixtures of local experts},
  journal = {Neural Computation},
  volume = {3},
  number = {1},
  pages = {79--87},
  year = {1991},
  doi = {10.1162/neco.1991.3.1.79}
}

@article{mitarai2018qcl,
  author = {Mitarai, K. and Negoro, M. and Kitagawa, M. and Fujii, K.},
  title = {{Quantum circuit learning}},
  journal = {Physical Review A},
  volume = {98},
  pages = {032309},
  year = {2018},
  doi = {10.1103/physreva.98.032309}
}

@article{wang2011rbf,
  author = {Wang, Ning and Liu, Min},
  title = {{Nuclear mass predictions with a radial basis function approach}},
  journal = {Physical Review C},
  volume = {84},
  pages = {051303},
  year = {2011},
  doi = {10.1103/physrevc.84.051303}
}

@article{neufcourt2019,
  author = {Neufcourt, L. and Cao, Y. and Nazarewicz, W. and Olsen, E. and Viens, F.},
  title = {Neutron drip line in the {Ca} region from {Bayesian} model averaging},
  journal = {Physical Review Letters},
  volume = {122}, pages = {062502}, year = {2019},
  doi = {10.1103/PhysRevLett.122.062502}
}

@article{dassie2020,
  author = {Dassie, A. C. and {Id Betan}, R. M.},
  title = {Estimate of the location of the neutron drip line for calcium isotopes from an exact {Hamiltonian} with continuum pair correlations},
  journal = {Physical Review C},
  volume = {102}, pages = {064301}, year = {2020},
  doi = {10.1103/PhysRevC.102.064301}
}

@article{zhou2024,
  author = {Zhou, Chang and Guo, Peng and Jiang, Xiaofei},
  title = {Giant Halo in {66Ca} Within Relativistic Continuum {Hartree--Bogoliubov} Theory Combined with {Lipkin--Nogami} Method},
  journal = {Particles},
  volume = {7}, number = {4}, pages = {1128--1138}, year = {2024},
  doi = {10.3390/particles7040069}
}

@article{cerezo2021vqa,
  author = {Cerezo, Marco and Arrasmith, Andrew and Babbush, Ryan and Benjamin, Simon C. and Endo, Suguru and Fujii, Keisuke and McClean, Jarrod R. and Mitarai, Kosuke and Yuan, Xiao and Cincio, Lukasz and Coles, Patrick J.},
  title = {Variational quantum algorithms},
  journal = {Nature Reviews Physics},
  volume = {3},
  pages = {625--644},
  year = {2021},
  doi = {10.1038/s42254-021-00348-9}
}

@article{schuld2021encoding,
  author = {Schuld, Maria and Sweke, Ryan and Meyer, Johannes Jakob},
  title = {{Effect of data encoding on the expressive power of variational quantum-machine-learning models}},
  journal = {Physical Review A},
  volume = {103},
  pages = {032430},
  year = {2021},
  doi = {10.1103/physreva.103.032430}
}

@article{utama2016bnn,
  author = {Utama, R. and Piekarewicz, J. and Prosper, H. B.},
  title = {{Nuclear mass predictions for the crustal composition of neutron stars: A Bayesian neural network approach}},
  journal = {Physical Review C},
  volume = {93},
  pages = {014311},
  year = {2016},
  doi = {10.1103/physrevc.93.014311}
}

@article{athanassopoulos2004,
  author = {Athanassopoulos, S. and Mavrommatis, E. and Gernoth, K. A. and Clark, J. W.},
  title = {Nuclear mass systematics using neural networks},
  journal = {Nuclear Physics A},
  volume = {743},
  pages = {222--235},
  year = {2004},
  doi = {10.1016/j.nuclphysa.2004.08.006},
}

@article{utama2017,
  author = {Utama, R. and Piekarewicz, J.},
  title = {Refining mass formulas for astrophysical applications: A {Bayesian} neural network approach},
  journal = {Physical Review C},
  volume = {96},
  pages = {044308},
  year = {2017},
  doi = {10.1103/PhysRevC.96.044308},
}

@article{niu2018,
  author = {Niu, Z. M. and Liang, H. Z.},
  title = {Nuclear mass predictions based on {Bayesian} neural network approach with pairing and shell effects},
  journal = {Physics Letters B},
  volume = {778},
  pages = {48--53},
  year = {2018},
  doi = {10.1016/j.physletb.2018.01.002},
}

@article{lovell2022,
  author = {Lovell, A. E. and Mohan, A. T. and Sprouse, T. M. and Mumpower, M. R.},
  title = {Nuclear masses learned from a probabilistic neural network},
  journal = {Physical Review C},
  volume = {106},
  pages = {014305},
  year = {2022},
  doi = {10.1103/PhysRevC.106.014305},
}

@article{mumpower2022,
  author = {Mumpower, M. R. and Sprouse, T. M. and Lovell, A. E. and Mohan, A. T.},
  title = {Physically interpretable machine learning for nuclear masses},
  journal = {Physical Review C},
  volume = {106},
  pages = {L021301},
  year = {2022},
  doi = {10.1103/PhysRevC.106.L021301},
}

@article{le2023,
  author = {Le, Xian-Kai and Wang, Nan and Jiang, Xiang},
  title = {Nuclear mass predictions with multi-hidden-layer feedforward neural network},
  journal = {Nuclear Physics A},
  volume = {1038},
  pages = {122707},
  year = {2023},
  doi = {10.1016/j.nuclphysa.2023.122707},
}

@article{li2020gsm,
  author = {Li, J. G. and Hu, B. S. and Wu, Q. and Gao, Y. and Dai, S. J. and Xu, F. R.},
  title = {Neutron-rich calcium isotopes within realistic {Gamow} shell model calculations with continuum coupling},
  journal = {Physical Review C},
  volume = {102},
  pages = {034302},
  year = {2020},
  doi = {10.1103/PhysRevC.102.034302},
}

@article{coraggio2020,
  author = {Coraggio, L. and {De Gregorio}, G. and Gargano, A. and Itaco, N. and Fukui, T. and Ma, Y. Z. and Xu, F. R.},
  title = {Shell-model study of calcium isotopes toward their drip line},
  journal = {Physical Review C},
  volume = {102},
  pages = {054326},
  year = {2020},
  doi = {10.1103/PhysRevC.102.054326},
}

@article{chen2023,
  author = {Chen, S. and Browne, F. and Doornenbal, P. and others},
  title = {Level structures of {$^{56,58}$Ca} cast doubt on a doubly magic {$^{60}$Ca}},
  journal = {Physics Letters B},
  volume = {843},
  pages = {138025},
  year = {2023},
  doi = {10.1016/j.physletb.2023.138025},
}

@article{yang2024dddf,
  author = {Yang, Zu-Xing and Fan, Xiao-Hua and Li, Zhi-Pan and Liang, Haozhao},
  title = {Data-driven density functional model for atomic nuclei},
  journal = {Physical Review C},
  volume = {109},
  number = {6},
  pages = {064312},
  year = {2024},
  doi = {10.1103/PhysRevC.109.064312}
}

@article{alam2025mirror,
  author = {Alam, N. A. and Qin, Z. Z. and Bao, M. and Zhang, K. Y. and Ma, C. and Lei, Y.},
  title = {Robust correlation between binding energies and charge radii of mirror nuclei},
  journal = {International Journal of Modern Physics E},
  volume = {34},
  number = {5},
  pages = {2550016},
  year = {2025},
  doi = {10.1142/S0218301325500168}
}

@article{zhang2026xgboost,
  author = {Zhang, X. Y. and Li, W. F. and Fang, J. Y.},
  title = {Improving nuclear mass predictions by correcting mass residuals using {eXtreme Gradient Boosting}},
  journal = {Chinese Physics C},
  volume = {50},
  number = {4},
  pages = {044101},
  year = {2026},
  doi = {10.1088/1674-1137/ae25cd}
}

@incollection{reinhard1991shf,
  author = {Reinhard, P.-G.},
  title = {The {Skyrme--Hartree--Fock} Model of the Nuclear Ground State},
  booktitle = {Computational Nuclear Physics 1: Nuclear Structure},
  editor = {Langanke, K. and Maruhn, J. A. and Koonin, S. E.},
  publisher = {Springer},
  address = {Berlin, Heidelberg},
  pages = {28--50},
  year = {1991},
  doi = {10.1007/978-3-642-76356-4_2}
}

@article{fedus2022switch,
  author = {Fedus, William and Zoph, Barret and Shazeer, Noam},
  title = {{Switch Transformers}: Scaling to Trillion Parameter Models with Simple and Efficient Sparsity},
  journal = {Journal of Machine Learning Research},
  volume = {23},
  number = {120},
  pages = {1--39},
  year = {2022},
  url = {https://jmlr.org/papers/v23/21-0998.html}
}

@inproceedings{lepikhin2021gshard,
  author = {Lepikhin, Dmitry and Lee, HyoukJoong and Xu, Yuanzhong and Chen, Dehao and Firat, Orhan and Huang, Yanping and Krikun, Maxim and Shazeer, Noam and Chen, Zhifeng},
  title = {{GShard}: Scaling Giant Models with Conditional Computation and Automatic Sharding},
  booktitle = {International Conference on Learning Representations},
  year = {2021},
  eprint = {2006.16668},
  archivePrefix = {arXiv},
  primaryClass = {cs.CL}
}

@misc{jiang2024mixtral,
  author = {Jiang, Albert Q. and Sablayrolles, Alexandre and Roux, Antoine and others},
  title = {{Mixtral} of Experts},
  year = {2024},
  eprint = {2401.04088},
  archivePrefix = {arXiv},
  primaryClass = {cs.LG}
}

@misc{dai2024deepseekmoe,
  author = {Dai, Damai and Deng, Chengqi and Zhao, Chenggang and others},
  title = {{DeepSeekMoE}: Towards Ultimate Expert Specialization in Mixture-of-Experts Language Models},
  year = {2024},
  eprint = {2401.06066},
  archivePrefix = {arXiv},
  primaryClass = {cs.CL}
}

@misc{yang2024qwen2,
  author = {Yang, An and Yang, Baosong and Hui, Binyuan and others},
  title = {{Qwen2} Technical Report},
  year = {2024},
  eprint = {2407.10671},
  archivePrefix = {arXiv},
  primaryClass = {cs.CL}
}

@misc{xai2024grok,
  author = {{xAI}},
  title = {Open Release of {Grok-1}},
  year = {2024},
  url = {https://x.ai/news/grok-os},
  note = {Released March 17, 2024}
}

@inproceedings{riquelme2021vision,
  author = {Riquelme, Carlos and Puigcerver, Joan and Mustafa, Basil and Neumann, Maxim and Jenatton, Rodolphe and Susano Pinto, Andr\'{e} and Keysers, Daniel and Houlsby, Neil},
  title = {Scaling Vision with Sparse Mixture of Experts},
  booktitle = {Advances in Neural Information Processing Systems},
  volume = {34},
  pages = {8583--8595},
  publisher = {Curran Associates, Inc.},
  year = {2021},
  url = {https://proceedings.neurips.cc/paper_files/paper/2021/hash/48237d9f2dea8c74c2a72126cf63d933-Abstract.html}
}

@article{schmidt2021mlgw,
  author = {Schmidt, Stefano and Breschi, Matteo and Gamba, Rossella and Pagano, Giulia and Rettegno, Piero and Riemenschneider, Gunnar and Bernuzzi, Sebastiano and Nagar, Alessandro and {Del Pozzo}, Walter},
  title = {Machine learning gravitational waves from binary black hole mergers},
  journal = {Physical Review D},
  volume = {103},
  number = {4},
  pages = {043020},
  year = {2021},
  doi = {10.1103/PhysRevD.103.043020}
}

@article{barenco1995gates,
  author = {Barenco, Adriano and Bennett, Charles H. and Cleve, Richard and DiVincenzo, David P. and Margolus, Norman and Shor, Peter and Sleator, Tycho and Smolin, John A. and Weinfurter, Harald},
  title = {Elementary gates for quantum computation},
  journal = {Physical Review A},
  volume = {52},
  number = {5},
  pages = {3457--3467},
  year = {1995},
  doi = {10.1103/PhysRevA.52.3457}
}

@inproceedings{paszke2019pytorch,
  author = {Paszke, Adam and Gross, Sam and Massa, Francisco and Lerer, Adam and Bradbury, James and Chanan, Gregory and Killeen, Trevor and Lin, Zeming and Gimelshein, Natalia and Antiga, Luca and Desmaison, Alban and Kopf, Andreas and Yang, Edward and DeVito, Zachary and Raison, Martin and Tejani, Alykhan and Chilamkurthy, Sasank and Steiner, Benoit and Fang, Lu and Bai, Junjie and Chintala, Soumith},
  title = {{PyTorch}: An Imperative Style, High-Performance Deep Learning Library},
  booktitle = {Advances in Neural Information Processing Systems},
  volume = {32},
  publisher = {Curran Associates, Inc.},
  year = {2019},
  url = {https://proceedings.neurips.cc/paper/2019/hash/bdbca288fee7f92f2bfa9f7012727740-Abstract.html},
}

@misc{bergholm2018pennylane,
  author = {Bergholm, Ville and Izaac, Josh and Schuld, Maria and Gogolin, Christian and others},
  title = {{PennyLane}: Automatic differentiation of hybrid quantum-classical computations},
  year = {2018},
  eprint = {1811.04968},
  archivePrefix = {arXiv},
  primaryClass = {quant-ph},
  doi = {10.48550/arXiv.1811.04968},
}

@article{buskirk2024shells,
  author = {Buskirk, Landon and Godbey, Kyle and Nazarewicz, Witold and Satu{\l}a, Wojciech},
  title = {Nucleonic shells and nuclear masses},
  journal = {Physical Review C},
  volume = {109},
  pages = {044311},
  year = {2024},
  doi = {10.1103/physrevc.109.044311},
}

@article{bender2006quadrupole,
  author = {Bender, M. and Bertsch, G. F. and Heenen, P.-H.},
  title = {Global study of quadrupole correlation effects},
  journal = {Physical Review C},
  volume = {73},
  pages = {034322},
  year = {2006},
  doi = {10.1103/physrevc.73.034322},
}

@article{satula1998staggering,
  author = {Satu{\l}a, W. and Dobaczewski, J. and Nazarewicz, W.},
  title = {Odd-Even Staggering of Nuclear Masses: Pairing or Shape Effect?},
  journal = {Physical Review Letters},
  volume = {81},
  pages = {3599--3602},
  year = {1998},
  doi = {10.1103/physrevlett.81.3599},
}

@article{hove2013massrelations,
  author = {Hove, D. and Jensen, A. S. and Riisager, K.},
  title = {Extrapolations of nuclear binding energies from new linear mass relations},
  journal = {Physical Review C},
  volume = {87},
  pages = {024319},
  year = {2013},
  doi = {10.1103/physrevc.87.024319},
}

@inproceedings{rahaman2019spectral,
  author = {Rahaman, Nasim and Baratin, Aristide and Arpit, Devansh and Draxler, Felix and Lin, Min and Hamprecht, Fred A. and Bengio, Yoshua and Courville, Aaron},
  title = {On the Spectral Bias of Neural Networks},
  booktitle = {Proceedings of the 36th International Conference on Machine Learning},
  series = {Proceedings of Machine Learning Research},
  volume = {97},
  pages = {5301--5310},
  publisher = {PMLR},
  year = {2019},
  url = {https://proceedings.mlr.press/v97/rahaman19a.html},
}

@article{moller2016frdm,
  author = {M{\"o}ller, P. and Sierk, A. J. and Ichikawa, T. and Sagawa, H.},
  title = {Nuclear ground-state masses and deformations: {FRDM}(2012)},
  journal = {Atomic Data and Nuclear Data Tables},
  volume = {109--110},
  pages = {1--204},
  year = {2016},
  doi = {10.1016/j.adt.2015.10.002},
}

@article{goriely2013hfb24,
  author = {Goriely, S. and Chamel, N. and Pearson, J. M.},
  title = {Further explorations of {Skyrme-Hartree-Fock-Bogoliubov} mass formulas. {XIII}. The 2012 atomic mass evaluation and the symmetry coefficient},
  journal = {Physical Review C},
  volume = {88},
  number = {2},
  pages = {024308},
  year = {2013},
  doi = {10.1103/PhysRevC.88.024308},
}

@article{kortelainen2014unedf2,
  author = {Kortelainen, M. and McDonnell, J. and Nazarewicz, W. and Olsen, E. and Reinhard, P.-G. and Sarich, J. and Schunck, N. and Wild, S. M. and Davesne, D. and Erler, J. and Pastore, A.},
  title = {Nuclear energy density optimization: Shell structure},
  journal = {Physical Review C},
  volume = {89},
  number = {5},
  pages = {054314},
  year = {2014},
  doi = {10.1103/PhysRevC.89.054314},
}

@article{baldo2013bcpm,
  author = {Baldo, M. and Robledo, L. M. and Schuck, P. and Vi{\~n}as, X.},
  title = {New {Kohn-Sham} density functional based on microscopic nuclear and neutron matter equations of state},
  journal = {Physical Review C},
  volume = {87},
  number = {6},
  pages = {064305},
  year = {2013},
  doi = {10.1103/PhysRevC.87.064305},
}

@article{goriely2009d1m,
  author = {Goriely, S. and Hilaire, S. and Girod, M. and P{\'e}ru, S.},
  title = {First {Gogny-Hartree-Fock-Bogoliubov} Nuclear Mass Model},
  journal = {Physical Review Letters},
  volume = {102},
  number = {24},
  pages = {242501},
  year = {2009},
  doi = {10.1103/PhysRevLett.102.242501},
}
\end{document}